# Space-charge-limited current density for nonplanar diodes with monoenergetic emission using Lie-point symmetries

N. R. Sree Harsha[1], Jacob M. Halpern[1], Adam M. Darr[1], and Allen L. Garner[1, 2, 3, a)]

[1]School of Nuclear Engineering, Purdue University, West Lafayette, IN 47906, USA

[2]Elmore Family School of Electrical and Computer Engineering, Purdue University, West Lafayette, IN 47907, USA

[3]Department of Agricultural and Biological Engineering, Purdue University, West Lafayette, IN 47907, USA

[a)]**Author to whom any correspondence should be addressed:** algarner@purdue.edu

*Abstract*: Understanding space-charge limited current density (SCLCD) is fundamentally and practically important for characterizing many high-power and high-current vacuum devices. Despite this, no analytic equations for SCLCD with nonzero monoenergetic initial velocity have been derived for nonplanar diodes from first principles. Obtaining analytic equations for SCLCD for nonplanar geometries is often complicated by the nonlinearity of the problem and over constrained boundary conditions. In this letter, we use the canonical coordinates obtained by identifying Lie-point symmetries to linearize the governing differential equations to derive SCLCD for any orthogonal diode. Using this method, we derive exact analytic equations for SCLCD with a monoenergetic injection velocity for one-dimensional cylindrical, spherical, tip-to-tip (t-t), and tip-to-plate (t-p) diodes. We specifically demonstrate that the correction factor from zero initial velocity to monoenergetic emission depends only on the initial kinetic and electric potential energies and not on the diode geometry and that SCLCD is universal when plotted as a function of the canonical gap size. We also show that SCLCD for a t-p diode is a factor of four larger than a t-t diode independent of injection velocity. The results reduce to previously derived results for zero initial velocity using variational calculus and conformal mapping.

Ordinary differential equations (ODEs) and partial differential equations (PDEs) are used extensively to model the behavior of many systems in applied physics and mathematics [1], including tumor growth [2, 3], heat transfer [4, 5], fluid mechanics [6], neutron transport [7], chemical engineering [8], electromagnetism [9], and electroporation [10-12]. Consider the general first-order ODE of the form

$$\frac{dy}{dx} = f(x, y), \tag{1}$$

where $f(x, y)$ is a smooth function of $x$ and $y$. If the function $f$ in (1) is a function of $x$ alone (i.e., $f(x, y) \equiv f(x)$), we can readily solve (1) by carrying out the quadrature

$$y = \int f(x)dx + c, \qquad (2)$$

for some integration constant $c$. If, however, $f(x,y)$ in (1) cannot be separated into a product of two functions such that $f(x,y) = g(x)h(y)$ for some smooth functions $g$ and $h$, the solution to the ODE cannot be obtained directly by carrying out a quadrature. This often arises for the applications described above [1-12], particularly for complicated geometries. For such situations, we can use the Lie-point symmetries of (1) to identify a set of canonical coordinates that reduces (1) to a form that can be solved directly by carrying out the quadrature (2) [13-15]. Similar techniques can be applied to solve higher-order ODEs and PDEs.

We demonstrate this approach in this letter by solving a second order ODE to derive exact analytic solutions for space-charge limited current density (SCLCD) in vacuum with nonzero electron injection velocity in any orthogonal diode geometry. The SCLCD is the maximum steady-state current density that can flow between a cathode and an anode in vacuum [16, 17]. Characterizing SCLCD is critical for numerous applications, including electric propulsion, nano vacuum transistors, thermionic energy converters, solar power conversion, high-power microwaves, and microplasma formation [16, 17]. For a one-dimensional (1-D) planar diode, the exact analytic equation for classical nonrelativistic SCLCD, first derived by Child and Langmuir [18, 19], is given by

$$J_{\mathrm{CL}} = \frac{\gamma V_g^{3/2}}{(x_A - x_C)^2}, \qquad (3)$$

where $\gamma = 4\epsilon_0 \sqrt{2e/m}/9$, $\epsilon_0$ is the permittivity of vacuum, $e$ and $m$ are the charge and mass of the electron, respectively, the anode is held at a constant voltage $V_g$ located at $x = x_A$, and the grounded cathode is located at $x = x_C$. Various studies have extended the CL law to account for multiple dimensions [20-27], nonplanar diode geometries [28-34], time-varying voltages [35-38], bipolar flow [39-42], relativistic effects [43, 44], trap-filled solids [45-47], and quantum effects [48-51].

When the electron injection velocity is zero, the electron charge density is infinite at the cathode and the corresponding SCLCD, given by (3), can only be obtained as the limit of an indeterminate 0/0 singularity at the cathode [52]. Jaffé avoided this singularity by deriving SCLCD for a constant nonzero electron injection velocity and recovered (3) by taking the limit as the ratio of the kinetic energy to electric potential energy approached zero [53]. Recently, Lafleur [54] validated Jaffé's results using particle-in-cell simulations and Huang et al. [55] used a sheet model to simulate the injection of electrons with either a monoenergetic or Maxwell-Boltzmann velocity distribution. In this letter, we extend (3) to include a constant nonzero initial velocity in a 1-D planar geometry before further generalizing SCLCD to *any* 1-D diode geometry represented by an orthogonal coordinate system, which we use to derive SCLCD for cylindrical, spherical, pin-to-pin, and pin-to-plate geometries.

We begin our analysis by considering a 1-D planar diode with cathode and anode, represented by infinite planes, located at $x = x_C$ and $x = x_A$, respectively. The anode is held at a constant potential $V_g$, while the cathode is grounded. Continuity implies that the local current density in planar coordinates is $J_p = \rho_p v_p$, where $\rho_p$ is charge density and $v_p$ is electron velocity across the gap. From conservation of energy in 1-D flow, given by $mv_p^2/2 = mv_0^2/2 + e\phi_p$ [54], we obtain

$$v_p = v_0 \sqrt{1 + \frac{2e\phi_p}{mv_0^2}}, \tag{4}$$

where $\phi_p$ is the electric potential in the planar gap and $v_0$ is the electron velocity at the cathode. Considering variation only in the *x*-direction, Poisson's equation in planar coordinates is given by

$$\frac{d^2\phi_p}{dx^2} = \frac{\rho_p}{\epsilon_0} = \frac{J_p}{\epsilon_0 v_p}. \tag{5}$$

Combining (4) and (5), we obtain

$$J_p = \epsilon_0 v_0 \frac{d^2\phi_p}{dx^2} \sqrt{1 + \frac{2e\phi_p}{mv_0^2}}. \tag{6}$$

Since continuity requires that $dJ_p/dx = 0$, planar SCLCD at the cathode, $J_{p,\text{SCL}}$, is given by

$$J_{p,\text{SCL}} = \lim_{x \to x_C}[\max(J_p)] = \max\left[\lim_{x \to x_C}(J_p)\right], \tag{7}$$

where we define max() as a functional that takes the expression on the right-hand side of (6) as an input and generates the maximum value of $J_p$ (denoted as $J_{p,\text{SCL}}$) as the output. We shall now present a general definition and derivation of the max() function.

***Derivation of the max() function.*** For generality, we consider the current density *J* in general canonical coordinates $\zeta$. We consider all current and electron velocity to be in the $\zeta$-direction. Writing (6) in canonical coordinates gives

$$J = \epsilon_0 v_0 \frac{d^2 \phi_\zeta}{d\zeta^2} \sqrt{1 + \frac{2e\phi_\zeta}{mv_0^2}} \tag{8}$$

with the grounded cathode at $\zeta = \zeta_C$ and the anode, biased to $V_g$, at $\zeta = \zeta_A$. To simplify the derivation, we define $v_D^2 = 2eV_g/m$ and the following dimensionless parameters [54]:

$$\bar{\phi}_\zeta = \frac{\phi_\zeta}{V_g}, \bar{\zeta} = \frac{\zeta - \zeta_C}{\zeta_A - \zeta_C}, \beta^2 = \frac{mv_0^2}{2eV_g}, \bar{J} = \frac{J_{\text{SCL}}(\zeta_A - \zeta_C)^2}{\epsilon_0 V_g v_D}. \tag{9}$$

Rewriting (8) in terms of the dimensionless parameters from (9), we obtain

$$\bar{J} = \frac{d^2 \bar{\phi}_\zeta}{d\bar{\zeta}^2} \sqrt{\bar{\phi}_\zeta + \beta^2}. \tag{10}$$

Multiplying both sides of (10) by $d\bar{\phi}/d\bar{\zeta}$ and simplifying yields

$$\frac{1}{2} \frac{d}{d\bar{\zeta}} \left( \frac{d\bar{\phi}_\zeta}{d\bar{\zeta}} \right)^2 = \frac{\bar{J}}{\sqrt{\beta^2 + \bar{\phi}_\zeta}} \left( \frac{d\bar{\phi}_\zeta}{d\bar{\zeta}} \right). \tag{11}$$

Multiplying both sides of (11) by $d\bar{\zeta}$ and integrating from 0 to $\bar{\zeta}$ gives a modified version of Poisson's equation (5) as

$$\frac{1}{2} \int_0^{\bar{\zeta}} d\left( \frac{d\bar{\phi}_\zeta}{d\bar{x}} \right)^2 = \bar{J} \int_0^{\bar{\phi}_\zeta} \frac{d\bar{\phi}_\zeta}{\sqrt{\beta^2 + \bar{\phi}_\zeta}}. \tag{12}$$

Simplifying (12), we obtain

$$\left( \frac{d\bar{\phi}_\zeta}{d\bar{\zeta}} \right)^2 = \left( \frac{d\bar{\phi}_\zeta}{d\bar{\zeta}} \right)^2 \bigg|_{\bar{\zeta} = 0} + 4\bar{J} \left( \sqrt{\beta^2 + \bar{\phi}_\zeta} - \beta \right). \tag{13}$$

Letting $\bar{\zeta}^*$ denote the location of the virtual cathode, where the electric field goes to zero, and $\bar{\phi}_\zeta^*$ denote the potential at $\bar{\zeta}^*$ gives

$$\left( \frac{d\bar{\phi}}{d\bar{\zeta}} \right)^2 \bigg|_{\bar{\zeta} = 0} = -4\bar{J} \left( \sqrt{\beta^2 + \bar{\phi}_\zeta^*} - \beta \right). \tag{14}$$

Substituting (14) into (13) and simplifying yields

$$\frac{d\bar{\phi}_\zeta}{d\bar{\zeta}} = \pm 2\sqrt{\bar{J}}\left(\sqrt{\beta^2 + \bar{\phi}_\zeta} - \sqrt{\beta^2 + \bar{\phi}_\zeta^*}\right)^{1/2}. \tag{15}$$

The normalized electric field, defined as $\bar{E} = -d\bar{\phi}_\zeta/d\bar{\zeta}$, is positive for $0 \leq \bar{\zeta} \leq \bar{\zeta}^*$ and negative for $\bar{\zeta}^* \leq \bar{\zeta} \leq 1$, while the electric field is zero at $\bar{\zeta} = \bar{\zeta}^*$. Integrating (15) while accounting for this sign change gives

$$-\int_0^{\bar{\phi}_\zeta^*} \frac{d\bar{\phi}_\zeta}{\left(\sqrt{\beta^2 + \bar{\phi}_\zeta} - \sqrt{\beta^2 + \bar{\phi}_\zeta^*}\right)^{1/2}} + \int_{\bar{\phi}_\zeta^*}^1 \frac{d\bar{\phi}_\zeta}{\left(\sqrt{\beta^2 + \bar{\phi}_\zeta} - \sqrt{\beta^2 + \bar{\phi}_\zeta^*}\right)^{1/2}}$$
$$= 2\sqrt{\bar{J}} \int_0^1 d\bar{\zeta}. \tag{16}$$

Defining $\eta_L = \sqrt{\beta^2 + 1}$ and $\eta^* = \sqrt{\beta^2 + \bar{\phi}_\zeta^*}$ allows to rewrite (16) as

$$\sqrt{\bar{J}} = \frac{2}{3}(\beta - \eta^*)^{1/2}(\beta + 2\eta^*) + \frac{2}{3}(\eta_L - \eta^*)^{1/2}(\eta_L + 2\eta^*). \tag{17}$$

We can find the extremum of $\bar{J}$ by setting $d\sqrt{\bar{J}}/d\eta^* = 0$ in (17), yielding $\eta_{\max}^* = \beta\eta_L/(\beta + \eta_L)$. Evaluating (17) at $\eta^* = \eta_{\max}^*$ gives the maximum stable current density in dimensionless parameters as

$$\bar{J}_{\text{SCL}} = \frac{4}{9}\left(\beta + \sqrt{1 + \beta^2}\right)^3. \tag{18}$$

Redimensionalizing $\bar{J}_{\text{SCL}}$ in (18) using (9) yields

$$J_{\text{SCL}} = \frac{\gamma V_g^{3/2}}{(\zeta_A - \zeta_C)^2}\left(\beta + \sqrt{1 + \beta^2}\right)^3. \tag{19}$$

Hence, we define the max() operator from (8) and (19) as

$$\max\left[\epsilon_0 v_0 \frac{d^2\phi_\zeta}{d\zeta^2}\sqrt{1 + \frac{2e\phi_\zeta}{mv_0^2}}\right] = \frac{\gamma V_g^{3/2}}{(\zeta_A - \zeta_C)^2}\left(\beta + \sqrt{1 + \beta^2}\right)^3. \tag{20}$$

We will next use this definition for a few orthogonal diode geometries. Returning to the 1-D planar problem, we substitute $J_p$ from (6) into (7) to obtain the SCLCD as

$$J_{p,\text{SCL}} = \left(\beta + \sqrt{1 + \beta^2}\right)^3 J_{\text{CL}}. \tag{21}$$

Poisson's equation (12) is separable in Cartesian coordinates since the spatial dependence can be isolated; however, this is not true for nonplanar diode geometries since the Laplacian in Poisson's equation becomes nonlinear. For such nonlinear ODEs, we can introduce canonical coordinates to make them separable and thereby derive analytical equations for SCLCD. To illustrate this approach, we first derive SCLCD in general orthogonal coordinates. We then consider 1-D concentric cylindrical, concentric spherical, tip-to-tip (t-t), and tip-to-plate (t-p) geometries as examples to further illustrate the method.

***General Coordinates.*** We can extend this technique to derive analytic expressions for SCLCD in any orthogonal diode geometry by first considering the generalized metric $ds^2 = (h_1 dq_1)^2 + (h_2 dq_2)^2 + (h_3 dq_3)^2 = \sum (h_i dq_i)^2$, where $ds$ represents the infinitesimal distance between any two points in the coordinate system, while the orthogonal coordinates and the scaling factors are represented by $q_i$ and $h_i$, respectively [56]. We shall only consider a 1-D diode geometry such that the potential $\phi_q$ varies only with $q_1$. The anode and cathode are at $q_A$ and the $q_C$, respectively. Continuity requires that the local current density $J_q = \rho_q v_q$, where $\rho_q$ and $v_q$ represent the electron charge density and velocity in the gap in the $q_1$-direction. We may write Poisson's equation as [57]

$$\frac{1}{h_1 h_2 h_3} \frac{d}{dq_1}\left(\frac{h_2 h_3}{h_1} \frac{d\phi_q}{dq_1}\right) = \frac{J_q}{\epsilon_0 v_q}. \tag{22}$$

Using conservation of energy for 1-D flow in general coordinates, given by $mv_q^2/2 = mv_0^2/2 + e\phi_q$, allows us to write (22) as

$$J_q = \epsilon_0 v_0 \left[\frac{1}{h_1 h_2 h_3}\frac{d}{dq_1}\left(\frac{h_2 h_3}{h_1}\frac{d\phi_q}{dq_1}\right)\right]\sqrt{1 + \frac{2e\phi_q}{mv_0^2}}, \tag{23}$$

where $v_0$ represents the velocity of the electrons at the cathode in the $q_1$-direction. To make (23) separable, we introduce the canonical coordinate $\zeta_q$ such that $d\zeta_q/dq_1 = h_1/(h_2 h_3)$. Using the canonical coordinate $\zeta_q$, defining $\phi_q\left(\zeta_q(q_1)\right) \equiv \phi_{\zeta_q}$, and applying the chain rule allows us to recast (23) as

$$J_q = \epsilon_0 v_0 \left[\frac{1}{(h_2 h_3)^2}\left(\frac{d^2\phi_{\zeta_q}}{d\zeta_q^2}\right)\right]\sqrt{1 + \frac{2e\phi_{\zeta_q}}{mv_0^2}}. \tag{24}$$

Continuity for 1-D flow in general orthogonal coordinates is given by $\nabla \cdot (\chi \vec{J}_q) = 0$, where $\chi = h_2 h_3$ for a flow in 1-D (along $q_1$), $\chi = h_3$ for a flow in two-dimensions (e.g., flow in the $q_1 q_2$-plane), and $\chi = 1$ for a flow in three-dimensions [58]. For a flow described in $q_1$, the continuity equation reduces to $d(h_2^2 h_3^2 J_q)/dq_1 = 0$. SCLCD $J_{q,SCL}$, which corresponds to the maximum current density emitted from the

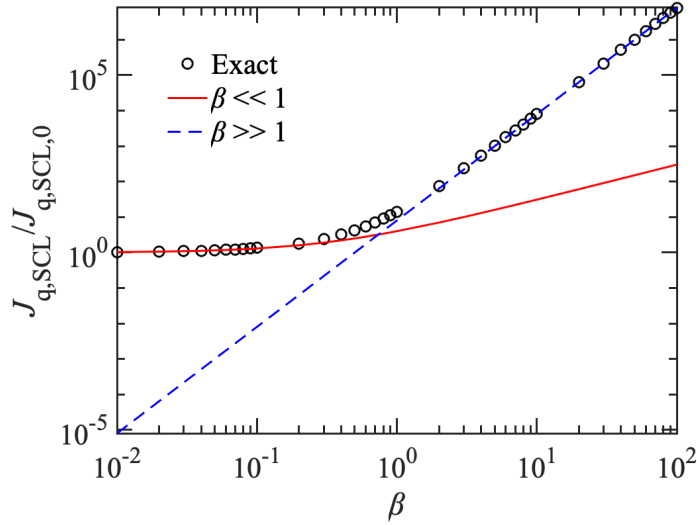

FIG. 1. The exact solution of the ratio of SCLCD in general coordinates to the corresponding SCLCD with zero injection velocity in general coordinates as a function of $\beta$ from (28) with limits for $\beta \ll 1$ from (30) and $\beta \gg 1$ from (31).

cathode, is given by

$$\lim_{q \to q_C}\left[\max((h_2 h_3)^2 J_q)\right] = \max\left[\lim_{q \to q_C}((h_2 h_3)^2 J_q)\right] = J_{q,\text{SCL}}[(h_2 h_3)^2|_{q=q_C}]. \tag{25}$$

We may then write SCLCD by substituting (24) into (25) to obtain

$$J_{q,\text{SCL}} = \frac{\epsilon_0 v_0}{(h_2 h_3)^2|_{q=q_C}} \max\left[\left(\frac{d^2 \phi_{\zeta_q}}{d\zeta_q^2}\right)\sqrt{1 + \frac{2e\phi_{\zeta_q}}{m v_0^2}}\right]. \tag{26}$$

Using (20) in (26) and substituting $\zeta_{q,A} = [\int h_1 dq/(h_2 h_3)]|_{q=q_A}$ and $\zeta_{q,C} = [\int h_1 dq/(h_2 h_3)]|_{q=q_C}$ gives SCLCD in the general orthogonal coordinate system as

$$J_{q,\text{SCL}} = \frac{(h_2 h_3)^{-2}|_{q=q_C}\, \gamma V_g^{3/2}\left(\beta + \sqrt{1+\beta^2}\right)^3}{\left(\left[\int \frac{h_1 dq}{h_2 h_3}\right]\big|_{q=q_A} - \left[\int \frac{h_1 dq}{h_2 h_3}\right]\big|_{q=q_C}\right)^2}. \tag{27}$$

Equation (27) represents the general form of the SCLCD by using the definition of metric $ds^2$ alone without having to solve the nonlinear Poisson's equation. Equation (27) is true for *any orthogonal geometry*. We may then write SCLCD in general orthogonal coordinates as

$$\frac{J_{q,\text{SCL}}}{J_{q,\text{SCL},0}} = \left(\beta + \sqrt{1+\beta^2}\right)^3, \tag{28}$$

where SCLCD with zero injection velocity $J_{q,\text{SCL},0}$ (i.e., $\beta = 0$) is given by

$$J_{q,\text{SCL},0} = \frac{(h_2 h_3)^{-2}|_{q=q_C}\, \gamma V_g^{3/2}}{\left(\left[\int \frac{h_1 dq}{h_2 h_3}\right]\Big|_{q=q_A} - \left[\int \frac{h_1 dq}{h_2 h_3}\right]\Big|_{q=q_C}\right)^2}. \tag{29}$$

For $\beta \ll 1$, we approximate (28) as

$$\frac{J_{q,\text{SCL}}}{J_{q,\text{SCL},0}} \approx 1 + 3\beta. \tag{30}$$

For $\beta \gg 1$, we approximate (28) as

$$\frac{J_{q,\text{SCL}}}{J_{q,\text{SCL},0}} \approx 8\beta^3. \tag{31}$$

Figure 1 shows $J_{q,\text{SCL}}/J_{q,\text{SCL},0}$ as a function of $\beta$ for the exact solution from (29) and the asymptotic solutions from (30) and (31). When $\beta = 0$, $J_{q,\text{SCL}} = J_{q,\text{SCL},0}$, as expected. We next apply this approach to common orthogonal coordinate systems.

*Example 1*. We first consider a concentric 1-D cylindrical diode in polar coordinates with the cathode located at $r = r_C$ and the anode located at $r = r_A$. With the metric in polar cylindrical coordinates given by $ds^2 = dr^2 + r^2 d\theta^2 + dz^2$, Poisson's equation for variation only in the $r$-direction may be written as

$$\frac{1}{r}\frac{d}{dr}\left(r\frac{d\phi_c}{dr}\right) = \frac{J_c}{\epsilon_0 v_c}, \tag{32}$$

where the local current density in cylindrical coordinates is given by $J_c = \rho_c v_c$, where $\rho_c$ and $v_c$ represent charge density and electron velocity in cylindrical coordinates. To make the ODE in (32) separable, we define a canonical coordinate $\zeta_c$ such that (32) may be reduced to a separable form [59]. Upon inspection, we may define $\zeta_c$ such that $d\zeta_c/dr = 1/r$, which yields $\zeta_c = \ln r$ and reduces the ODE in (32) to

$$J_c = \frac{\epsilon_0 v_0}{r^2}\left(\frac{d^2 \phi_{\zeta_c}}{d\zeta_c^2}\right)\sqrt{1 + \frac{2e\phi_{\zeta_c}}{mv_0^2}}, \tag{33}$$

where $\phi_c(\zeta_c(r)) \equiv \phi_{\zeta_c}$. Continuity for 1-D flow in polar coordinates requires that $d(r^2 J_c)/dr = 0$ [58, 60], which implies that the quantity $r^2 J_c$ is independent of $r$. Hence, the maximum current density (SCLCD) at the cathode is given by

$$\lim_{r \to r_C}[\max(r^2 J_c)] = \max\left[\lim_{r \to r_C}(r^2 J_c)\right] = r_C^2 J_{c,\text{SCL}}. \tag{34}$$

Maximizing (33) by substituting it in (34) yields

$$r_C^2 J_{c,\text{SCL}} = \epsilon_0 v_0 \max\left[\left(\frac{d^2\phi_{\zeta_c}}{d\zeta_c^2}\right)\sqrt{1 + \frac{2e\phi_{\zeta_c}}{mv_0^2}}\right]. \tag{35}$$

Using (20) in (35) and noting that $\zeta_{c,A} = \ln r_A$ and $\zeta_{c,C} = \ln r_C$, we obtain

$$J_{c,\text{SCL}} = \frac{\gamma V_g^{3/2}}{r_C^2(\ln r_A - \ln r_C)^2}\left(\beta + \sqrt{1+\beta^2}\right)^3. \tag{36}$$

For $\beta = 0$ (i.e., zero injection velocity), (36) reduces to $J_{c,\text{SCL}} = \gamma V_g^{3/2}/[r_C^2(\ln r_A - \ln r_C)^2]$, which agrees with the result from variational calculus (VC) [30] and conformal mapping (CM) [31]. Note that this derivation is completely independent of the VC and CM techniques. Fig. 2(a) shows $J_{c,\text{SCL}}$ as a function of the canonical gap size $\delta_c = r_C^2[\ln(r_C/r_A)]^2$ for various $\beta$.

***Example 2***. We next consider a concentric 1-D spherical diode with the anode and cathode located at $r_A$ and $r_C$, respectively. Defining $ds^2 = dr^2 + r^2 d\phi^2 + r^2 \sin^2\phi\, d\theta^2$ as the metric in spherical coordinates and assuming variation only in $r$-direction gives Poisson's equation as

$$\frac{1}{r^2}\frac{d}{dr}\left(r^2\frac{d\phi_s}{dr}\right) = \frac{J_s}{\epsilon_0 v_s}, \tag{37}$$

where the local current density in spherical coordinates is given by $J_s = \rho_s v_s$. We define the canonical coordinate $\zeta_s$ such that $d\zeta_s/dr = 1/r^2$, which yields $\zeta_s = -1/r$. Rewriting (37) in terms of canonical coordinates and defining $\phi_s(\zeta_s(r)) \equiv \phi_{\zeta_s}$, we obtain

$$J_s = \frac{\epsilon_0 v_0}{r^4}\left(\frac{d^2\phi_{\zeta_s}}{d\zeta_s^2}\right)\sqrt{1 + \frac{2e\phi_{\zeta_s}}{mv_0^2}}. \tag{38}$$

Continuity for 1-D flow in spherical coordinates requires that $d(r^4 J_s)/dr = 0$ [58, 60], which implies that the quantity $(r^4 J_s)$ is independent of $r$. Hence, SCLCD at the cathode can be obtained by

$$\lim_{r \to r_C}[\max(r^4 j_s)] = \max\left[\lim_{r \to r_C}(r^4 j_s)\right] = r_C^4 J_{s,\text{SCL}}. \tag{39}$$

Maximizing (38) by substituting it in (39) yields

$$r_C^4 J_{s,\text{SCL}} = \epsilon_0 v_0 \max\left[\left(\frac{d^2\phi_{\zeta_s}}{d\zeta_s^2}\right)\sqrt{1 + \frac{2e\phi_{\zeta_s}}{mv_0^2}}\right]. \tag{40}$$

Using (20) in (40) and noting that $\zeta_{s,A} = -1/r_A$ and $\zeta_{s,C} = -1/r_C$, we obtain

$$J_{s,SCL} = \frac{\gamma V_g^{3/2} r_A^2}{r_C^2 (r_A - r_C)^2} \left(\beta + \sqrt{1+\beta^2}\right)^3. \quad (41)$$

For zero injection velocity (i.e., $\beta = 0$), (41) reduces to $J_{s,SCL} = \gamma V_g^{3/2} r_A^2 / [r_C^2(r_A - r_C)^2]$, which agrees with VC [30] and provides an independent verification of this approach. Figure 2(b) shows $J_{s,SCL}$ as a function of the canonical gap size $\delta_s = r_C^2 [r_A - r_C]^2 / r_A^2$ for $\beta$. Note that SCLCD is independent of geometry when plotted as a function of the canonical gap size, so one may consider this plot as universal and use the appropriate metric to convert from canonical to physical gap size.

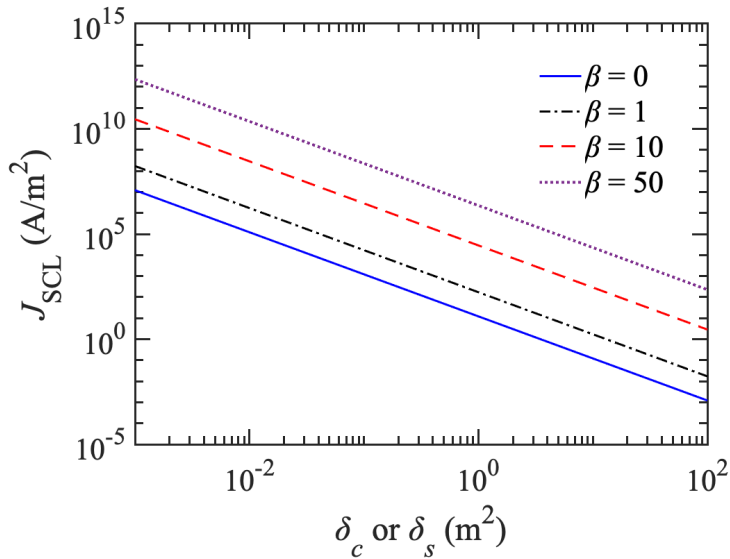

FIG. 2. SCLCD ($J_{SCL}$) as a function of canonical gap size $\delta_c = r_C^2 [\ln(r_C/r_A)]^2$ or $\delta_s = r_C^2 [r_A - r_C]^2 / r_A^2$ for cylindrical or spherical coordinates, respectively, for various $\beta$. The applied voltage is $V_g = 30$ kV. Note that $J_{SCL}$ is the same for any geometry for a given canonical gap size.

***Example 3.*** We next consider tip-to-tip (t-t) and tip-to-plate (t-p) geometries, where the cathode and the anode are represented by hyperboloids $\eta_C$ and $\eta_A$, respectively, in prolate spheroidal coordinates [32]. If $J_{t-t}$ represents the local current density in the t-t geometry and the metric in prolate spheroidal coordinates is given by $ds^2 = a^2[(\sinh^2 \xi + \sin^2 \eta)(d\xi^2 + d\eta^2) + (\sinh^2 \xi \sin^2 \eta)d\varphi^2]$, Poisson's equation, assuming variation only in the $\eta$-direction, is given by [32]

$$\frac{1}{a^2} \frac{1}{(\sinh^2 \xi + \sin^2 \eta) \sin \eta} \frac{d}{d\eta}\left(\sin \eta \frac{d\phi_\eta}{d\eta}\right) = \frac{J_{t-t}}{\epsilon_0 v_\eta}. \quad (42)$$

Using conservation of energy, given by $mv_\eta^2/2 = mv_0^2/2 + e\phi_\eta$, and defining the canonical coordinate $\zeta_t$ such that $d\zeta_t/d\eta = 1/\sin \eta$ and $\phi_\eta(\zeta_t(\eta)) \equiv \phi_{\zeta_t}$ reduces (42) to

$$J_{t-t} = \frac{\epsilon_0 v_0}{a^2(\sinh^2\xi + \sin^2\eta)\sin^2\eta}\left(\frac{d^2\phi_{\zeta_t}}{d\zeta_t^2}\right)\sqrt{1 + \frac{2e\phi_{\zeta_t}}{mv_0^2}}. \quad (43)$$

Continuity for 1-D flow in prolate spheroidal coordinates requires that $d(h_2^2 h_3^2 J_{t-t})/d\eta = d((\sinh^2\xi + \sin^2\eta)\sin^2\eta\, J_{t-t})/d\eta = 0$, where $h_2^2 = a^2(\sinh^2\xi + \sin^2\eta)$ and $h_3^2 = a^2 \sinh^2\xi \sin^2\eta$ [58]. At the tip of the cathode represented by $\xi = 0$, we note that $\sinh\xi = 0$ and the continuity equation reduces to $d(a^2 \sin^4\eta\, J_{t-t})/d\eta = 0$. Hence, SCLCD at the cathode tip can be written as

$$\lim_{\eta\to\eta_C}[\max(a^2\sin^4\eta\, J_{t-t})] = \max\left[\lim_{\eta\to\eta_C}(a^2\sin^4\eta\, J_{t-t})\right] = J_{t-t,\text{SCL}}\, a^2\sin^4\eta_C. \quad (44)$$

Maximizing (43) by substituting it into (44) and noting that $\sinh\xi = 0$ at the cathode tip yields

$$J_{t-t,\text{SCL}}\, a^2\sin^4\eta_C = \epsilon_0 v_0 \max\left[\left(\frac{d^2\phi_{\zeta_t}}{d\zeta_t^2}\right)\sqrt{1+\frac{2e\phi_{\zeta_t}}{mv_0^2}}\right]. \quad (45)$$

Using (20) in (45) and noting that $\zeta_{t,A} = \ln[\tan(\eta_A/2)]$ and $\zeta_{t,C} = \ln[\tan(\eta_C/2)]$ gives SCLCD in t-t geometry at the cathode tip as

$$J_{t-t,\text{SCL}} = \frac{\gamma V_g^{3/2}}{a^2 \sin^4\eta_C \left(\ln\left[\tan\left(\frac{\eta_A}{2}\right)\right] - \ln\left[\tan\left(\frac{\eta_C}{2}\right)\right]\right)^2}\left(\beta + \sqrt{1+\beta^2}\right)^3. \quad (46)$$

When the electron injection velocity goes to zero ($\beta = 0$), (46) reduces to $J_{t-t,\text{SCL}} = \gamma V_g^{3/2} a^{-2} \sin^{-4}\eta_C (\ln[\tan(\eta_A/2)] - \ln[\tan(\eta_C/2)])^{-2}$, which agrees with VC and CM [32].

Assuming identical tips (i.e., $\eta_A = \pi - \eta_C$) and defining the distance between the apex of the tips as $D_0$ gives $a^2 = D_0(D_0 + R)$, where $R$ is the radius of the tips [32]. Next, defining $\mu = D_0/R = \cot^2\eta_A = \cot^2\eta_C$ yields $\cos^2(\eta_A) = \cos^2(\eta_C) = D_0/(D_0 + R)$ and $\sin^4\eta_C = \sin^4\eta_A = R^2(D_0 + R)^{-2}$. Using these definitions and further noting that $x_A - x_C = D_0$ allows us to write (46) in terms of $J_{\text{CL}}$ with $D_0^2 = (x_A - x_C)^2$ as

$$\frac{J_{t-t,\text{SCL}}}{J_{\text{CL}}} = \frac{\mu(\mu+1)}{4[\ln(\sqrt{1+\mu}+\sqrt{\mu})]^2}\left(\beta + \sqrt{1+\beta^2}\right)^3, \quad (47)$$

where we used $\sqrt{1+\mu} - \sqrt{\mu} = \left(\sqrt{1+\mu}+\sqrt{\mu}\right)^{-1}$ to obtain the denominator.

The SCLCD in t-p geometries can be obtained from (46) by noting that the anode (a plate) is represented in prolate spheroidal coordinates with $\eta_A \to \pi/2$ in (35a), which necessitates $\ln[\tan(\eta_A/$ 2)] $\to 0$ [32]. Applying this and the limit of zero injection velocity to (46) gives $J_{\text{t-p,SCL}} = \gamma V_g^{3/2} a^{-2} \sin^{-4}\eta_C (\ln[\tan(\eta_C/2)])^{-2}$, which agrees with VC and CM [32]. Hence, SCLCD near the cathode tip in t-p geometry is given by

$$\frac{J_{\text{t-p,SCL}}}{J_{\text{CL}}} = \frac{\mu(\mu+1)}{[\ln(\sqrt{1+\mu}+\sqrt{\mu})]^2}\left(\beta+\sqrt{1+\beta^2}\right)^3. \tag{48}$$

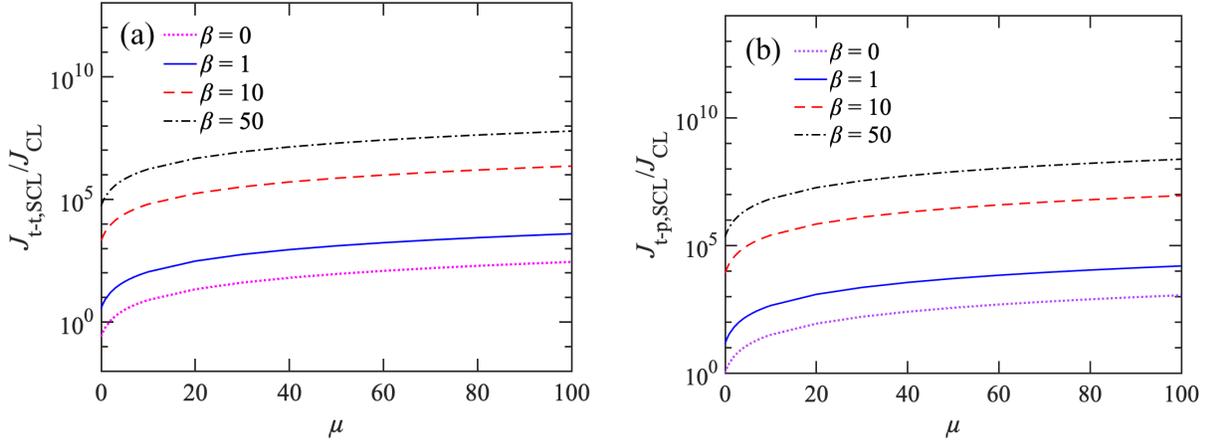

Fig. 3. SCLCD normalized to CL as a function of $\mu = D_0/R$ for various $\beta$ for (a) tip-tip ($J_{\text{t-t,SCL}}/J_{\text{CL}}$) and (b) tip-to-plate ($J_{\text{t-p,SCL}}/J_{\text{CL}}$) geometries.

Writing (47) and (48) in this form further shows that $J_{\text{t-p,SCL}} = 4 J_{\text{t-t,SCL}}$, which was not apparent from the forms we derived for zero injection velocity using CM [32]. Figure 3 shows $J_{\text{t-t,SCL}}/J_{\text{CL}}$ and $J_{\text{t-p,SCL}}/J_{\text{CL}}$ as a function of $\mu$ for various $\beta$. Equations (47) and (48) show that the current density increases by a factor of four as the anode tip becomes a horizontal plate for a given injection velocity.

In summary, we have demonstrated how to obtain canonical coordinates to simplify the ODE in Poisson's equation to make it separable to solve for SCLCD in general orthogonal coordinates for nonzero monoenergetic injections velocities and then for four simple geometries. Of particular note, we have provided independent validation of previous calculations of SCLCD using VC for spherical, tip-to-tip, and tip-to-plate geometries. We have also shown that the correction factor to account for initial velocity for any coordinate system is independent of geometry, so it is straightforward to write SCLCD for monoenergetic emission once the SCLCD with zero injection velocity for a 1-D geometry is known. This may be valuable for complicated geometries that may not be amendable to this analysis, such as those we have studied previously using CM [31]. We have also demonstrated that 1-D SCLCD is independent of geometry once written as a function of canonical gap size (cf. Fig. 2); therefore, one may obtain SCLCD for a given geometry by applying the appropriate metric to the universal SCLCD. For ODEs and PDEs for which the canonical coordinates are not obvious, one may use Lie-point symmetries to derive the corresponding canonical coordinates to simplify the ODEs and PDEs [1, 59]. This method provides various potential directions for future research. For instance, Lie-point symmetries can simplify the governing PDEs in cylindrical crossed-field diode geometries [61]. This method can also be used to simplify the ODEs and PDEs in tumor growth [2, 3], heat transfer [4, 5], fluid mechanics [6], neutron transport [7], chemical engineering [8], electromagnetism [9], and electroporation [10-12], facilitating the derivation of theories for more realistic geometries.

Finally, we also note that other current definitions [62, 63] are claimed to be SCLCD for monoenergetic injection of electrons, but more rigorous analysis has shown that they correspond to the bifurcation point in the diode [64]. SCLCD, as defined by the maximum current density permissible in the diode, is not related to the onset of reflection of particles but to the inherent nonlinear behavior of governing ODEs [64]. Ongoing studies will apply VC and CM to simplify the Poisson's equation and delineate how the SCLCD and bifurcation point are obtained using the appropriate differential equations, boundary conditions, and functional for extremization [65]. Future work will also apply these techniques to the transition across electron emission mechanisms, including field emission and quantum SCLCD [66], and to time-dependent conditions [67].

This material is based upon work supported by the Air Force Office of Scientific Research under Award No. FA9550-19-1-0101. J.M.H. gratefully acknowledges funding from an Undergraduate Research Scholarship from the Purdue School of Nuclear Engineering. A.M.D. gratefully acknowledges funding from a Purdue Doctoral Fellowship. We also gratefully acknowledge L. K. Ang and John Luginsland for useful discussions.

## AUTHOR DECLARATIONS

### Conflict of Interest

The authors have no conflict of interest to disclose.

### Data Availability

Data sharing is not applicable to this article as no new data were created or analyzed in this study.